\documentclass[a4paper]{jpconf}
\usepackage{graphicx}
\begin{document}
\title{Scalar dark matter in an extra dimension inspired model}

\author{Roberto Lineros$^1$, Fabio Pereira dos Santos$^{2,3}$}

\address{$^1$Instituto de F\'{\i}sica Corpuscular -- CSIC/U. Valencia, Spain\\
$^2$Centro Federal de Educa\c c\~ao Tecnol\'ogica Celso Suckow da Fonseca, Petr\'opolis, RJ, Brazil
\\ $^3$Departamento de F\'isica, Pontif\'icia Universidade Cat\'olica do Rio de Janeiro, RJ, Brazil}

\ead{rlineros@ific.uv.es, fabio.alex@fis.puc-rio.br}

\begin{abstract}
In this work we consider a singlet scalar propagating in a flat large extra dimension. The first Kaluza-Klein mode associated to this singlet scalar will be a viable dark matter candidate.
The tower of new particles enriches the calculation of the relic density due effect of coannihilation. For large mass splitting, the model converges to the predictions of the singlet dark matter model. For nearly degenerate mass spectrum, coannihilations increase the cross-sections used for direct and indirect dark matter searches. We investigate the impact of the Kaluza-Klein tower associated to singlet scalar for indirect and direct detection of dark matter.
\end{abstract}

\section{Introduction}
Although the relic density of dark matter can be determined with high precision, a very important question still remains without answer, what the nature of dark matter is and also which theory is able to explain it considering all recent observations.
In this work we propose a model based on flat extra dimension that is different from the universal extra dimension model in the field propagation~\cite{Lineros:2014jba}. We assume that all the standard model particles are confined to a 3-brane and only a singlet scalar (and gravity) can propagate in the $3+\delta$ space. The stability of the first nonzero state of the KK tower associated to the singlet scalar is ensured by a remnant U(1)-like symmetry that arises after integrating over the extra dimension. We calculate the annihilation cross section and spin independent cross section including coannihilation as a function of dark matter mass. Also, we show that annihilation and spin independent cross sections of dark matter undergo an increase when we include coannihilation. We include in our results for annihilation cross section bounds from FERMI-LAT~\cite{Ackermann:2013yva}, MAGIC~\cite{Aleksic:2013xea}, and HESS~\cite{Abramowski:2011hc} and for spin independent cross section bounds from XENON100~\cite{Aprile:2012nq}, LUX~\cite{Akerib:2013tjd}, CDMSlite~\cite{Agnese:2013jaa}, CRESST~\cite{Angloher:2014myn}, and DAMA/LIBRA~\cite{Savage:2008er}.

\section{The model}
In this work we consider a model based on a modified version of the large extra dimension~\cite{ArkaniHamed:1998rs}. For our purpose the scalar sector is composed by the the SM higgs doublet H, and by a real scalar $S$, which is the only field in the model that depends on the extra dimension $y$ that is compactified in a circle with radius $R$. The lagrangian of the scalar sector is divided into three parts~\cite{Lineros:2014jba}:
\begin{eqnarray}
	\mathcal{L}_{\rm scalar}(y) &=& \mathcal{L}_{H} + \mathcal{L}_{S}(y)  + \mathcal{L}_{HS}(y) \ ,
\end{eqnarray}
where the lagrangian for the $\mathcal{L}_{H}, \mathcal{L}_{S}(y)$ and $\mathcal{L}_{HS}(y)$ are:
\begin{eqnarray}
\mathcal{L}_{H} \propto (D_{\mu} H)^{\dagger} D^{\mu} H - \frac{m_D^2}{2 v^2} \left(H^{\dagger}H - \frac{v^2}{2}\right)^2 \ ,\\
\mathcal{L}_{S}(y) = \frac{1}{2} \partial_{N} S(y) \partial^{N} S(y) - \frac{m_{S}^2}{2} S(y)^2 - \mu S - \frac{\lambda_{3}}{3} S(y)^3 - \frac{\lambda_4}{4} S(y)^4 \ ,\\
\mathcal{L}_{HS}(y) = - \lambda_{H}\left(H^{\dagger}H - \frac{v^2}{2}\right)S(y) - \lambda_{2H} \left(H^{\dagger}H - \frac{v^2}{2}\right)S(y)^2 \ ,
\end{eqnarray}
where in the unitary gauge: $H^{T} = (0 , (v + h))/\sqrt{2}$. The higgs vacuum expectation value is $v\simeq 246$ GeV and $m_D$ the mass term.\\
Due to the circular shape of the extra dimension,
the periodic boundary condition for the extra coordinate can be expressed as
$S(y)=S(y+2\pi R)$. Consequently, the field
$S(y)$ appears in a set of Fourier modes,
\begin{equation}
S(y) = \frac{1}{\sqrt{\pi R M_5}} \left\{ \frac{S_0}{\sqrt{2}} + \sum_{n=1}^{\infty} S_n \cos{\left(\frac{n y}{R}\right)} + \sum_{n=1}^{\infty} P_n \sin{\left(\frac{n y}{R}\right)} \right\} \, ,
\end{equation}
where $S_n$ and $P_n$ correspond to the KK modes of $S(y)$. In order to obtain the effective lagrangean of the model we integrate over the extra dimension $y$, the result is:
\begin{eqnarray}
\label{eq:lseff1}
\mathcal{L}_S^{\rm eff} &=& \frac{1}{2}\left( \partial_{\mu} S_0 \partial^{\mu} S_0 - m_S^2 S_0^2 \right) + \frac{1}{2}\left( \sum_{n=1}^{\infty} \partial_{\mu} S_n \partial^{\mu} S_n - (m_S^2 + \frac{n^2}{R^2}) S_n^2 \right) \nonumber \\
& & + \frac{1}{2}\left( \sum_{n=1}^{\infty} \partial_{\mu} P_n \partial^{\mu} P_n - (m_S^2 + \frac{n^2}{R^2}) P_n^2 \right) - \frac{\omega_3}{3} \mathcal{V}_3 - \frac{\omega_4}{4} \mathcal{V}_4 \, . 
\end{eqnarray}

The KK modes, $S_n$ and $P_n$, are degenerate in mass as we can realize from equation \ref{eq:lseff1}. Hence, we can define
\begin{equation}
\chi_n = \frac{S_n + i P_n}{\sqrt{2}} \  {\rm and} \ \chi_n^{*} = \frac{S_n - i P_n}{\sqrt{2}} \, ,
\end{equation}
it reduces the equation~\ref{eq:lseff1} to:
\begin{equation}
\mathcal{L}_S^{\rm eff} = \frac{1}{2}\left( \partial_{\mu} S_0 \partial^{\mu} S_0 - m_S^2 S_0^2 \right) + \sum_{n=1}^{\infty} \partial_{\mu} \chi_n^{*} \partial^{\mu} \chi_n - m_{S_n}^2 \chi_n^{*}\chi_n - \frac{\omega_3}{3} \mathcal{V}_3 - \frac{\omega_4}{4} \mathcal{V}_4 \, .
\end{equation}

The interaction terms $\mathcal{V}_3$ and $\mathcal{V}_4$ are, for example:
\begin{eqnarray}
\mathcal{V}_3 &=& S_0^3 + 3 S_0 \left(|\chi_1|^{2} + |\chi_2|^{2}\right)  + 3 \left(\chi_1^2 \chi_2^{*} + \chi_2 {\chi_1^{*}}^2  \right) \, ,
\end{eqnarray}

\begin{eqnarray}
\mathcal{V}_4 &=& S_0^4 + 12 S_0^2 \left(|\chi_1|^2 + |\chi_2|^2\right) + 12 S_0 \left(\chi_1^2 \chi_2^{*} + \chi_2 {\chi_1^{*}}^2\right) \nonumber \\ 
&&+ 6 \left(|\chi_1|^4 + 4 |\chi_1|^2 |\chi_2|^2 + |\chi_2|^4\right) \, ,
\end{eqnarray}
We can observe that a global $U(1)$ symmetry with charges $Q(\chi_n) = n$ is present in the lagrangian.
So, the lightest KK state, $\chi_1$, is stable and therefore it is the dark matter candidate.\\

\section{Relic abundance and coannihilations}
The relic density of a dark matter particle is determined by its self-annihilation cross section and also by annihilation processes involving heavier particles. We consider that it was in thermal equilibrium in the early universe and also decoupled when it was non relativistic. In order to reproduce the correct dark matter relic density we have to solve the Boltzmann equation
\begin{eqnarray}
\frac{dn_{\chi}}{dt} + 3Hn_{\chi}=-\langle \sigma_{\mbox{\tiny{eff}}} v\rangle\left(n^2_{\chi}-n^2_{eq}\right),
\end{eqnarray} 
where $H$ is the Hubble expansion rate, $n_{eq}$ is the number density at thermal equilibrium and $\langle \sigma_{\mbox{\tiny{eff}}} v\rangle$ is the thermally averaged effective cross section for annihilation times the relative velocity. The effective cross section for our model is given by:
\begin{eqnarray}
\label{eq:seff}
\sigma_{\rm c} = \sigma_{0} \frac{\displaystyle \sum_{i=1}^n  (1+\Delta_i)^{3} e^{-2 x \Delta_i}}{\displaystyle \left[\sum_{i=1}^n (1+\Delta_i)^{3/2}e^{-x\Delta_i} \right]^2} = \sigma_0 f(x,\Delta_2,n) \, ,
\end{eqnarray}
where the mass splitting between the DM and the {\it i}-th KK particles can be written  in terms of $\Delta_2$:
\begin{equation}
\Delta_i = \sqrt{\displaystyle 1 + \frac{(i^2 - 1)}{3} \Delta_2 (\Delta_2 + 2)} - 1 \, ,
\end{equation}
where $\Delta_2$ ranges between 0 and 1. In this work we parametrize the mass degeneracy as:
\begin{equation}
\label{eq:xi}
\xi_{KK} = \frac{2}{\pi} \tan{\left( \frac{\pi}{2} \Delta_{2}\right)}.
\end{equation}

In order to be consistent with indirect detection searches that give $\langle \sigma v \rangle \sim 10^{-25} \, {\rm cm}^3/{\rm s}$ we find the mass splitting $\xi_{KK} > 10^{-3}$ or in terms of number of KK modes $n\sim 10$. The parameter $\xi_{KK}$ shows regions where DM phenomenology is affected by coannihilations (see~\cite{Lineros:2014jba} for details).

\section{Results and discussions}
 In order to compute the theoretical prediction of the relic density of our model we use public software tools LanHEP~\cite{Semenov:2008jy}  and MicrOmegas~\cite{Belanger:2013oya}. 

\begin{figure}[ht]
\centering
\includegraphics[scale=2.2]{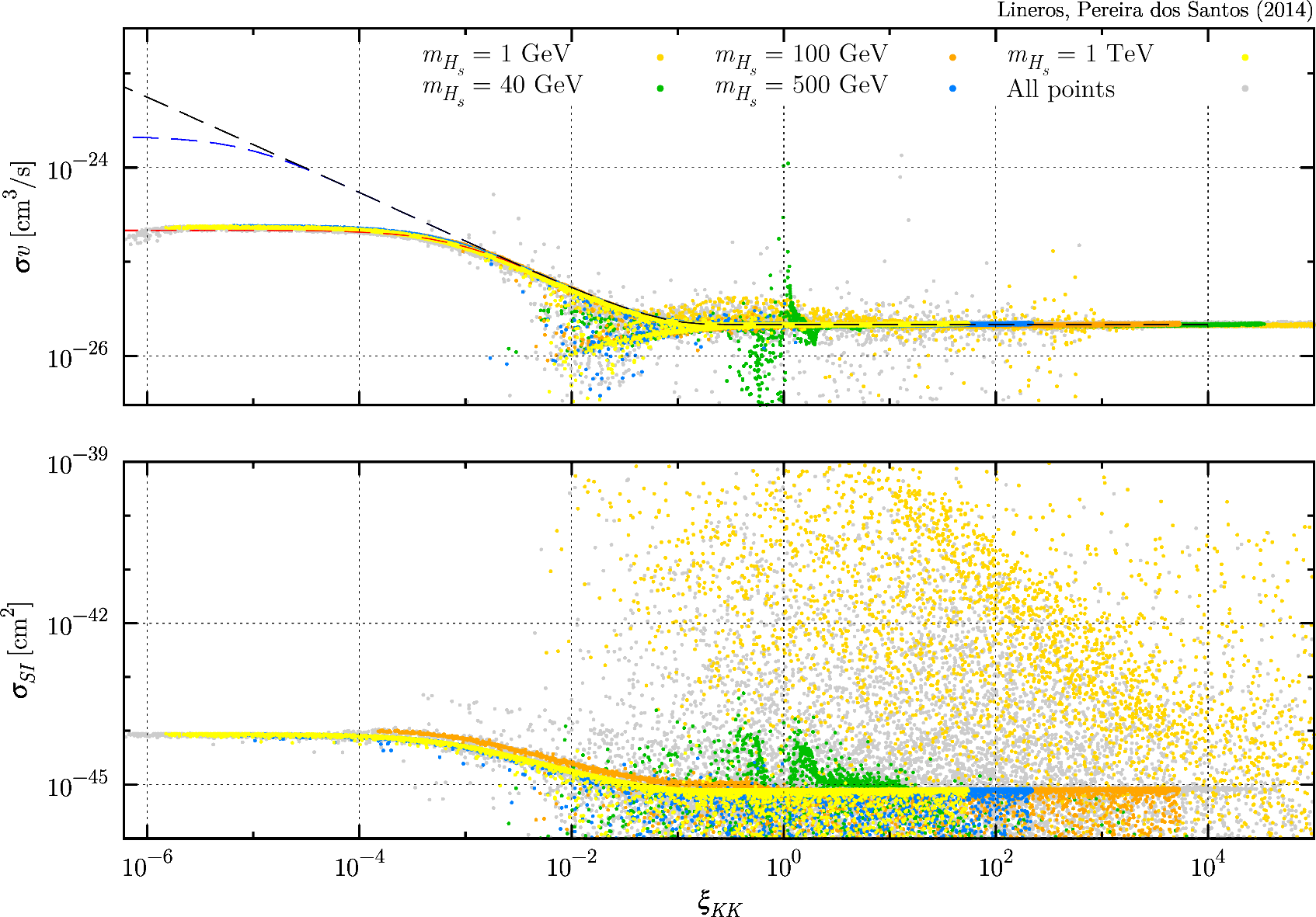}   
\caption{Annihilation cross section (top) and Spin Independent cross section (bottom) versus $\xi_{KK}$. Both cross sections converge to a fixed value for low values of $\xi_{KK}$.}
\label{sled_mass}
\end{figure}

\begin{figure}[ht!]
\centering
\includegraphics[scale=2.2]{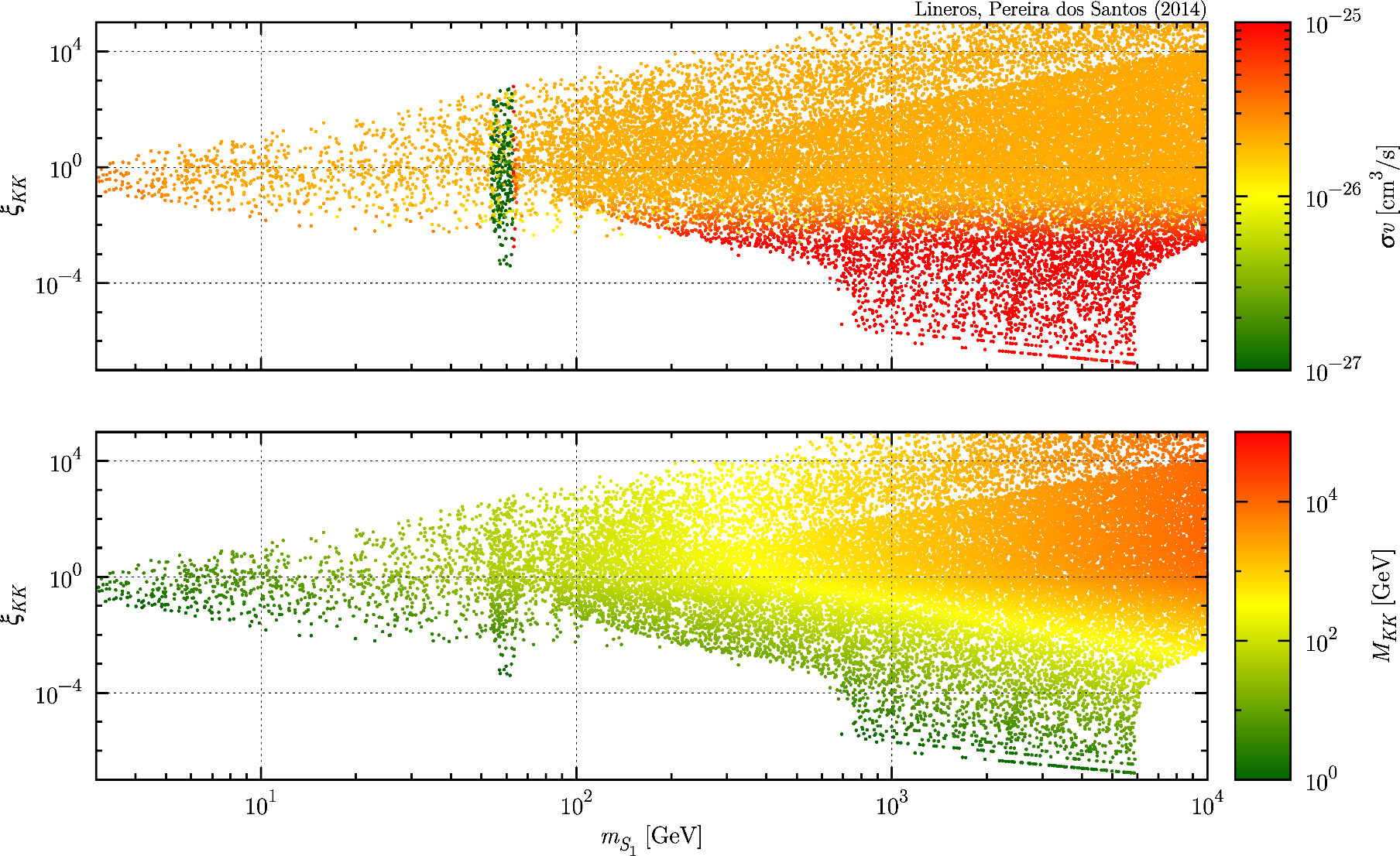}   
\caption{Top-panel: $\xi_{KK}$ versus DM mass. Color scale corresponds to the annihilation cross section value. Bottom-panel: $\xi_{KK}$ versus DM mass. Color scale corresponds to the value of $M_{KK}$.}
\label{sled_mass_pan}
\end{figure}

Figure \ref{sled_mass} shows the mass splitting dependence of the annihilation cross section and the spin independent cross section in terms of mass degeneracy parameter $\xi_{KK}$ . For lower values of $\xi_{KK}$ we observe a similar behavior in both cross sections  . For $\xi_{KK} < 10^{-2}$, i.e. mass splitting of the order of $1\%$ of the DM mass, the cross sections converge to common values: $\sigma_{SI} \simeq 10^{-44} \mbox{cm}^2$ and $\sigma v\simeq 2 \times 10^{-25}\mbox{cm}^3/s$. In fact, this effect is associated to coannihilations. From figure \ref{sled_mass_pan} we observe that points with $\sigma v > 3\times 10^{-26} \mbox{cm}^3/s$ are in the mass range $150 \mbox{Gev}< m_{S_1}< 10 \mbox{TeV}$.


\begin{figure}[ht!]
\centering
\includegraphics[scale=2.2]{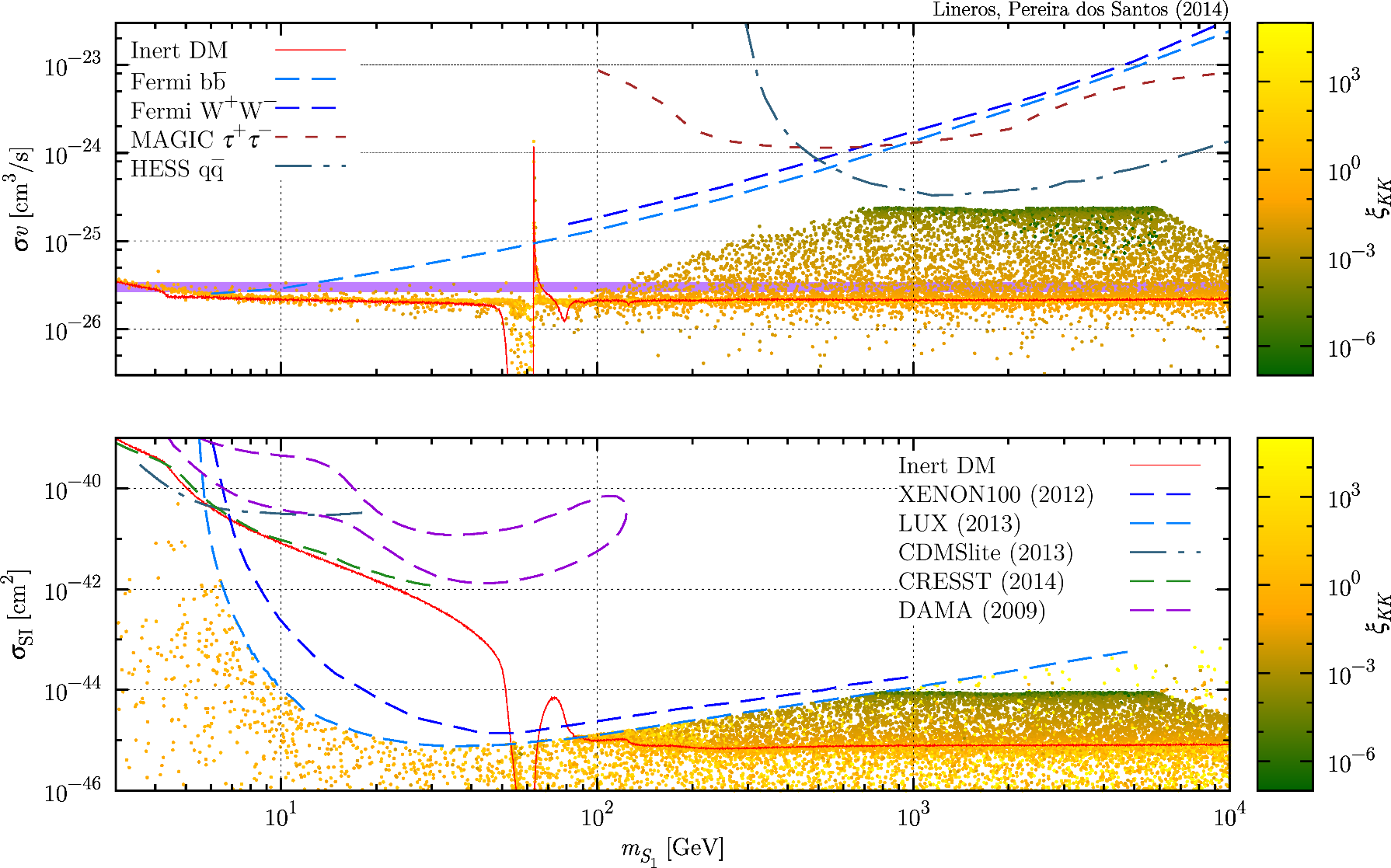}
\caption{Annihilation cross section (top-panel) and Spin Independent cross section (bottom-panel) vs DM mass excluding points that do not satisfy latest LUX bounds~\cite{Akerib:2013tjd}.}
\label{sled_mass_lux}
\end{figure}

In the top panel of figure \ref{sled_mass_lux} we show the annihilation cross section as a function of dark matter mass $m_{S_1}$. The solid red line correponds to the predictions of the inert singlet DM model . Bounds from FERMI-LAT~\cite{Ackermann:2013yva}, MAGIC~\cite{Aleksic:2013xea}, and HESS~\cite{Abramowski:2011hc} are displayed in dashed lines. The color scale correponds to the value of the degeneracy parameter $\xi_{KK}$. Smaller values of $\xi_{KK}$ lead to enhanced annihilations due to more Kaluza Klein modes were involved in the early universe.
In the bottom panel we show the spin independent cross section versus the dark matter mass $m_{S_1}$. Bounds from  XENON100~\cite{Aprile:2012nq}, LUX~\cite{Akerib:2013tjd}, CDMSlite~\cite{Agnese:2013jaa}, CRESST~\cite{Angloher:2014myn}, and DAMA/LIBRA~\cite{Savage:2008er} are also shown.
The color bar indicates the value of $\xi_{KK}$ where for lower values there is an enhancement of both cross sections due to coannihilations.
On the other hand, when $\xi_{KK}$ is large most of the points fall close to the predictions of the inert singlet DM model.

\section{Conclusions} 
The model we present in this work allows us to study the effect of coannihilations when several particles are quasi degenerate in mass. We found that for a degenerate mass spectrum both annihilation and spin independent cross section converge to fixed values  $\sigma_{SI} \simeq 10^{-44} \mbox{cm}^2$ and $\sigma v\simeq 2 \times 10^{-25} \mbox{cm}^3/s$. This enhancement can be translated to a mass splitting among the KK particles smaller than $100$ GeV. For larger mass splittings, the model converges to the predictions of the inert scalar DM model.

\section{Acknowledgments}
This work was supported by the Spanish MINECO under grants FPA2014-58183-P, and MULTIDARK CSD2009-00064 (Consolider-Ingenio 2010 Programme); by Generalitat Valenciana grant PROMETEOII/2014/084, and Centro de Excelencia Severo Ochoa SEV-2014-0398 and by Brazilian agencies Capes and CNPq. R.L. is supported by a Juan de la Cierva contract JCI-2012-12901 (MINECO). We would like to thank you the organizers of the TAUP 2015 for an interesting conference.

\section*{References}



\end{document}